# ASSESSMENT OF POLLUTANT SPREAD FROM A BUILDING BASEMENT WITH THREE VENTILATION SYSTEMS

## JUSLIN KOFFI

**LEPTIAB – University of La Rochelle, FRANCE**

## INTRODUCTION

Ventilation aims at providing a sufficient air renewal for ensuring a good indoor air quality (IAQ), yet building energy policies are leading to adapting various ventilation strategies minimising energy losses through air renewal. A recent IAQ evaluation campaign in French dwellings shows important pollution of living spaces by VOCs such as formaldehyde, acetaldehyde or hexanal, particularly in buildings equipped with a garage [1]. Besides, radon emission from soil is a subject of concern in many countries. Several studies are done to understand its release mode and deal with the spread of this carcinogen gas [2]. This paper aims to experimentally assess a contaminant spread from a house basement using mechanical exhaust and balanced ventilation systems, and natural ventilation.

## METHODS

The investigations are carried out in the experimental house MARIA [3] by means of tracer gas methods [4]. MARIA consists of four bedrooms, a bathroom and a shower at first-level, a living-room, a kitchen and toilets at ground-level, all linked by a hall; the basement is connected to the hall by a closed door. The pollution source is simulated by a release of hexafluoride sulphur ($SF_6$) in the basement at 2 ml/s flow rate for 5 hours; concentrations are measured there and in five rooms of the habitable levels.

## RESULTS

The measured $SF_6$ concentrations are presented by Figure 1 for exhaust ventilation: concentration in the basement reaches 690 mg/m$^3$. Pollution levels in the exhaust rooms are quite comparable (up to 180 mg/m$^3$), likewise in the main rooms (bedrooms and living, about 55 mg/m$^3$). The maximum concentration in the hall is around 230 mg/m$^3$. The other fraction of the pollutant is probably eliminated through the building envelope air leakage; for achieving this, wind and temperature difference between indoors and outdoors play the key role. That high dispersion of the contaminant within the dwelling tends to partly provide an explanation of the IAQ survey results mentioned above and an idea of what could be exposure to radon release in single-family houses.

Comparable results are found with natural ventilation. However, bedroom 3 is more polluted than the bathroom where the concentration is 150 mg/m$^3$: this is due to the low driving forces of natural ventilation during the studies. In addition, concentrations at ground-level are inferior to 14 mg/m$^3$.

For these exhaust ventilation systems, the depressurisation produced in the habitable rooms by the exhaust fan and the natural driving forces engenders great airflow from the basement towards these levels of the dwelling. This leads to the pollution of all experienced rooms.

The best performance is achieved with balanced ventilation (see Figure 2). Only a little pollution is measured in the habitable levels of the dwelling. The maximum level does not exceed 14 mg/m$^3$ in the hall and the exhaust rooms, and concentration in the main rooms are lower than 2 mg/m$^3$: further studies [5] show a suitable drainage of air and pollutant flows from the second room category to the first one with balanced ventilation.

Here, the tracer gas accumulates more in the injection room (930 mg/m$^3$) due to a lower air change rate (0.21 ach vs. 0.41 ach for exhaust ventilation). But in fact, the equilibrium between air extraction and air supply results in an almost zero-pressure zone in the hall, so that airflow from the basement towards the hall is very low. This is a characteristic of balanced ventilation systems. The use of such a system seems to bring a better protection against contaminants from basement.

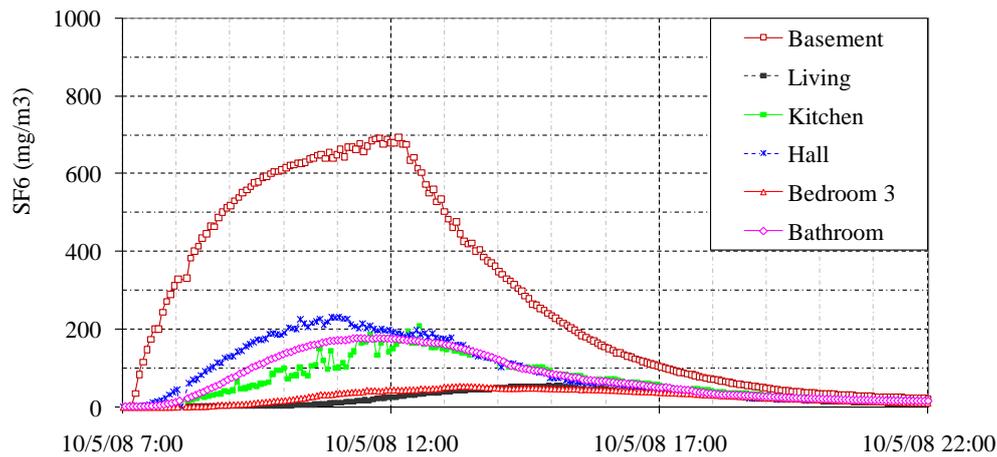

**Figure (1), Measured concentrations for mechanical exhaust ventilation system.**

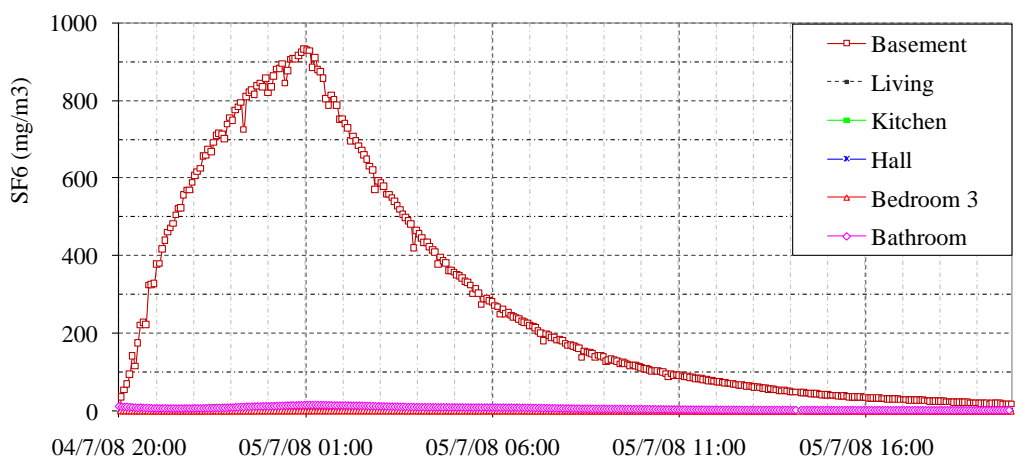

**Figure (2), Measured concentrations for mechanical balanced ventilation system.**

All these results should be interpreted by taking into account the building configuration and envelope air leakage as well as the test conditions. However, they give an evaluation of basement pollutants elimination by ventilation systems.

#### ACKNOWLEDGMENT

This study was carried out as part of a Ph.D. thesis at CSTB Research Centre in Marne-la-Vallée (Paris). It was directed by Pr. Francis Allard (LEPTIAB – University of La Rochelle, France) and Dr. Jean-Jacques Akoua (CSTB, France).